\documentclass{webofc}

\usepackage[varg]{txfonts}   
\woctitle{XIII$^{\mathrm{th}}$ International Conference on Meson-Nucleon Physics and the Structure of the Nucleon}
\begin{document}
\title{Physics perspectives at JLab with a polarized positron beam}
%
%

\author{Eric Voutier\inst{1}\fnsep\thanks{\email{voutier@lpsc.in2p3.fr}}}

\institute{Laboratoire de Physique Subatomique et de Cosmologie \\
           LPSC, Universit\'e Grenoble-Alpes, CNRS/IN2P3 , 53 avenue des Martyrs, 38026 Grenoble, France
          }

\abstract{%
  Polarized positron beams are in some respect mandatory complements to polarized electron beams. 
  The advent of the PEPPo concept for polarized positron production opens the possibility for the developement at the 
  Jefferson Laboratory of a continuous polarized positron beam. The benefits of such a beam for hadronic structure studies 
  are discussed, together with the technical and technological challenges to face.
}
\maketitle

\section{Introduction}
\label{intro}

Symmetry properties in physics, either verified or violated, are the basis of a quantum theory 
and fundamental tools of experimental investigations. For instance: the CPT symmetry is considered 
as a basic and mandatory property of any physics law; within the standard model, the violation of 
the electroweak symmetry is associated to massive gauge bosons via the Higgs mechanism~\cite{Hig64}; 
also, it is the spontaneous breaking of the chiral symmetry which confers to pions their very singular 
role in nuclear physics. On the experimental side, the violation of the parity symmetry in polarized 
electron scattering became over the past decade a privilegied tool for the investigation of the 
strange content of the nucleon~\cite{{Maa05},{Arm05},{Ani06}} and develops today towards tests of 
the standard model~\cite{You07}. As part of these fundamental experimental tools, the comparison 
between polarized electron and polarized positron scatterings takes advantage of the lepton beam charge 
to reveal and isolate peculiar features of physics.


\section{Physics motivations}
\label{sec-PhyMot}

The benefits of polarized positron beams complementary to polarized electron beams are numerous and 
extend from the precise study of the structure of hadrons to tests of the standard model and the 
exploration of new Physics~\cite{JPos09}. We discuss hereafter two specific examples of particular 
interest to the Physics research program at the Jefferson Laboratory (JLab). 

\subsection{Nucleon electromagnetic form factors}
\label{sub-EFF}

Following the very first measurements of polarization transfer observables in elastic electron ($e$) 
scattering off protons ($p$), the accuracy of the Born approximation for the description of the 
electromagnetic $ep$ interaction was questionned. The possible importance of higher orders in the 
$\alpha$-development of the electromagnetic current was suggested as an hypothesis to reconciliate 
cross section and polarization transfer experimental data~\cite{Gui03}. The electromagnetic shape of 
the nucleon would consequently be represented by the 3 generalized complex form factors
\begin{equation}
\widetilde G_{M} = G_{M} + e_b \, \delta {\widetilde G}_M \,\,\,\,\,\,\, \widetilde G_{E} = G_{E} + e_b 
\, \delta {\widetilde G}_E \,\,\,\,\,\,\, \widetilde F_{3} = \delta {\widetilde F}_3 
\end{equation}
where $e_b$ represents the lepton beam charge. These expressions involved 8 unknown quantities that should 
be recovered from experiments~\cite{Rek04}. Considering unpolarized leptons, the reduced experimental 
cross section ($\sigma_R$) writes 
\begin{equation}
\sigma_R = G^2_{M} + \frac{\epsilon}{\tau} \, G^2_{E} 
- 2 e_b \, G_{M} \, \Re{\left[ f_0 \left( \delta {\widetilde G}_M , \delta {\widetilde F}_3 \right) \right]} 
- 2 e_b \, \frac{\epsilon}{\tau} \, G_{E} \, \Re{\left[ f_1 \left( \delta {\widetilde G}_E , \delta {\widetilde F}_3 
\right) \right]}
\end{equation}
involving 5 unknwon quantities. The transverse ($P_t$) and longitudinal ($P_l$) polarization of the nucleon 
provide 2 additionnal and different linear combinations of the same quantities 
\begin{eqnarray}
 \sigma_R \, P_{t} & = & - \, P_b \, \sqrt{\frac{2\epsilon(1-\epsilon)}{\tau}} 
 \left( G_{E} G_{M} - e_b \, G_E \Re \left[ \delta {\widetilde G}_M \right] - e_b \, 
 G_{M} \Re{ \left[ f_0 \left( \delta {\widetilde G}_M , \delta {\widetilde F}_3 \right) \right]} \right) \\
 \sigma_R P_{l} & = & P_b \, \sqrt{1-\epsilon^2} \left( G_{M}^2 - 2e_b \, 
 G_{M} \Re \left[  f_2 \left( \delta {\widetilde G}_M , \delta {\widetilde F}_3 \right) \right] \right) \, .
\end{eqnarray}
where $P_b$ is the lepton beam polarization. Therefore, measuring cross sections and nucleon polarization observables 
in polarized electron and positron scatterings off the nucleon allows a model independent determination of the nucleon 
electromagnetic form factors. 

\subsection{Nucleon generalized parton distributions}
\label{sub-GPD}

Generalized parton distributions (GPDs) encode the intimate structure of hadronic matter in terms of quarks 
and gluons and are linked via sum rules to electromagnetic form factors (see \cite{Die03} for a review). 
They are involved in any deep process and are preferentially accessed in hard lepto-production of real 
photons i.e. deeply virtual Compton scattering (DVCS). This process competes with the known Bethe-Heitler 
(BH) reaction where real photons are emitted from initial or final leptons instead of the probed hadronic 
state. The lepton beam charge and polarization dependence of the cross section off protons writes~\cite{Die09}
\begin{equation}
\sigma_{{P_b}0}^{e_b} = \sigma_{BH} + \sigma_{DVCS} + P_b \, 
\widetilde{\sigma}_{DVCS} + e_b \, \sigma_{INT} + P_b e_b \, \widetilde{\sigma}_{INT}
\end{equation}
where the index $INT$ denotes the interference between the BH and DVCS amplitudes. Polarized electron 
scattering provide the reaction amplitude combinations
\begin{equation}
\sigma_{OO}^{-} = \sigma_{BH} + \sigma_{DVCS} - \sigma_{INT} \,\,\,\,\,\,\, \sigma_{+O}^{-} - \sigma_{-O}^{-} 
= 2 \widetilde{\sigma}_{DVCS} - 2 \widetilde{\sigma}_{INT} 
\end{equation}
and the comparison between polarized electrons and positrons provide the additional combinations
\begin{equation}
\sigma_{OO}^{+} - \sigma_{OO}^{-} = 2 \sigma_{INT} \,\,\,\,\,\,\, 
\left[ \sigma_{+O}^{+} - \sigma_{+O}^{-} \right] - \left[ \sigma_{-O}^{+} - \sigma_{-O}^{-} \right] 
= 4 \widetilde{\sigma}_{INT} \, .
\end{equation}
Consequently, measuring real photon lepto-production off the nucleon with opposite charge polarized leptons allows to 
isolate the 4 unknown contributions to the cross section involving GPDs. The separation of nucleon GPDs requires 
additionnal polarization observables measured with polarized targets ($S$). The full lepton beam charge and polarizations 
dependence of the cross section writes~\cite{Die09}  
\begin{equation}
\sigma_{{P_b} S}^{e_b} = \sigma_{{P_b} 0}^{e_b} + S \left[ P_b \, \Delta\sigma_{BH} + ( \Delta\widetilde{\sigma}_{DVCS} 
+ P_b \, \Delta\sigma_{DVCS} ) + e_b \, ( \Delta\widetilde{\sigma}_{INT} + P_b \, \Delta\sigma_{INT} ) \right] 
\end{equation}
where once again, polarized electron and positron observables allow to separate the new 4 unknown contributions to the 
cross section off polarized targets. Polarized positron beams then appear as a mandatory complement to polarized 
electron beams for a model independent determination of nucleon GPDs.


\section{Polarized positron production at JLab}
\label{sec-PosPro}

At the next generation of non storage ring facilities, polarized positrons are expected to be obtained from the 
creation of $e^+e^-$ pairs by circularly polarized photons produced either from the Compton backscattering of a 
polarized laser light on a few-GeV electron beam~\cite{Omo06}, or the synchrotron radiation of a multi-GeV electron 
beam travelling through a helical undulator~\cite{Ale08}. A new approach~\cite{Gra11} is developed at JLab relying 
on the bremsstrahlung production of circularly polarized photons from polarized electrons. As opposed to other 
schemes, the PEPPo (Polarized Electrons for Polarized Positrons) concept is potentially capable of operating at 
lower beam energies (a few-MeV) with high polarization transfer from incident electrons to created 
positrons~\cite{Kur10}, and would open access to low/moderate polarized positron beam intensities for a wide 
Community ranging from Material Science to Nuclear and Hadronic Physics.

The PEPPo experiment~\cite{Gra11} at the injector of the Continuous Electron Beam Accelerator Facility (CEBAF) 
of JLab has been designed to evaluate and demonstrate the PEPPo concept for a polarized positron source. A few-$\mu$A 
highly polarized (85\%) continuous electron beam of 8.25~MeV/$c$ is transported to a tungsten production target 
(0.1-1.0~mm) where polarized positrons are produced. The collection and momentum selection of these positrons 
is insured by a strong solenoid lens followed by $\pm$90$^\circ$ bend $\mathrm{D} \overline{\mathrm{D}}$ 
spectrometer dipoles. A second solenoid at the exit of the spectrometer is optimizing the positron transport to a 
Compton transmission polarimeter where the positron polarization is inferred from the measurement of the absorption of 
polarized bremsstrahlung generated polarized photons in a 7.5~cm long polarized iron target. The polarimeter has been 
characterized~\cite{Ade13} with the well-known CEBAF electron beam, and the positron momentum dependence of the 
experimental asymmetry has been measured. The PEPPo Collaboration reported~\cite{{Vou13},{Gra13}} significant 
on-line experimental asymmetries supporting high positron polarizations, and indicated at the current stage of 
the analysis a lower limit of 50\% for the positron polarization.


\section{Perspectives and challenges}
\label{sec-JLabPer}

PEPPo clearly opens the ability to produce polarized positrons with any polarized electron beam driver, as long 
as the beam kinetic energy exceeds the pair production threshold. At low energies, suitable beam currents for 
Physics would be obtained from the mutual optimization of (among others) the initial electron beam intensity 
and the production target thickness~\cite{Dum11}, while high energy production may look more favorable in 
these respects. The optimum capture of the produced positrons for further acceleration (Nuclear and Hadronic Physics) or 
decceleration (Material Science Physics) is an additional issue which may require supplementary systems like  
cooling- and/or damping-rings and/or cavities.

In the context of the JLab accelerator facility, three different locations can potentially host a polarized 
positron source: the CEBAF injector ($\sim$10~MeV and $\sim$100~MeV), the Free Electron Laser ($\sim$100~MeV), 
and the north linac end (up to 12 GeV). While each location has specific issues, the common technical and technological 
challenges comprise: a high intensity polarized positron source capable of a few-mA continuous beam delivery with 
high polarization degree; a high power (10-100 kW) production target challenging for heat dissipation, 
radiation management, and accelerator integration; an optimized positron collection system for the delivery of quality 
beams for physics experiments; a positron beam \textit{shaping} system to prepare for further acceleration or decelaration. Optimization of each scenario is obviously a multi-parameter problem but conceptual studies are already promising: a 
full lattice simulation based on a 100~MeV/1~mA electron beam driver showed that positron beam intensities up to 300~nA 
can be expected with a momentum spread of 10$^{-2}$ and a beam emittance $(\varepsilon_x \varepsilon_y) = (1.6,1.7)$~mm.mrad~\cite{Gol10}, PEPPo indicating further that polarizations larger than 60\% of the initial electron beam polarization can be achieved. Such a beam would definitely fulfills requirements for the study of the nucleon structure, 
for instance with the CLAS12 detector. 


\section{Conclusion}
\label{sec-Conc}

The merits of polarized and/or unpolarized positron beams for the Physics program at JLab is comparable to the 
benefits of polarized with respect to unpolarized electrons. Such capabilities would shed light on the 
2$\gamma$-physics and would enable a model-independent experimental investigation of the nucleon structure down 
to the partonic scale. It would also benefit Nuclear Physics studies at low energies and potentially Material 
Science Physics at very low to thermal energies. 

Significant R\&D efforts are necessary to address several technical and technological challenges among which 
high power absorbers and positron beam capture directly impact performances of beam delivery for Physics. 
The simulation results of specific conceptual studies and the experimental results of the PEPPo experiment are   
very encouraging and supporting further developements towards a polarized positron beam at JLab.



\end{document}